\documentclass[%
 reprint,
 amsmath,amssymb,
 aps,
 pre
]{revtex4-2}

\usepackage{graphicx}
\usepackage{dcolumn}
\usepackage{bm}
\usepackage{microtype}
\usepackage{hyperref}
\usepackage{comment}
\usepackage{longtable} 
\usepackage{booktabs}
\usepackage{bbm}
\usepackage{xcolor}

\definecolor{prlblue}{rgb}{0.18,0.19,0.57}

\hypersetup{
    colorlinks=true,       
    linkcolor=prlblue,     
    citecolor=prlblue,     
    urlcolor=prlblue,      
    pdfstartview={FitH},   
    bookmarksopen=true     
}

\newcommand{\dd}{\mathrm{d}}
\newcommand{\dOm}{\mathrm{d}\Omega}
\newcommand{\dA}{\mathrm{d}A}
\newcommand{\bu}{\bm{u}}
\newcommand{\bx}{\bm{x}}
\newcommand{\by}{\bm{y}}

\begin{document}

\preprint{APS/123-QED}

\title{Asphericity lifts the degeneracy among ordering mechanisms on curved surfaces}

\author{Anže Božič}
\thanks{anze.bozic@ijs.si}
\affiliation{%
Department of Theoretical Physics, Jožef Stefan Institute, SI-1000 Ljubljana, Slovenia
}


\begin{abstract}
Spherical surfaces conceal the one-body signature of particle organization: area-uniform, conductor, and curvature-weighted measures coincide there, so the ordering mechanism leaves no trace in the density. Hard spheres and Coulomb charges confined to the sphere thus produce identical mean densities, and distinctions between their particle arrangements appear only in many-body correlations. Aspherical deformation lifts this degeneracy, and we show that on an ellipsoid, different ordering mechanisms separate by virtue of a single exponent $p$ of the distance from the shell center to its tangent plane. To leading order in deformation, the characteristic exponent $p$ can be obtained both from the quadrupole of the spherical structure factor and from the radius of gyration. Two interaction families demonstrate how this can be realized in practice: the Riesz $s$-energy, whose limiting measure is determined by the power $s$ and exhibits a transition at $s=d=2$, and the screened Coulomb potential, whose measure and thus the characteristic exponent are tuned continuously by the salt concentration. The value of the characteristic exponent $p$ positions the arrangement of particles between the conductor and packing limits, and its response to a control parameter such as salt concentration can be used experimentally to identify the interaction family.
\end{abstract}

\maketitle

\section*{Introduction}

The physical principles governing self-organization of point-like particles on closed curved surfaces underpin diverse phenomena, from protein arrangements on viral capsids~\cite{caspar,zandi,rochal2024proteinaceous} to colloidal stabilization on Pickering emulsion droplets~\cite{bausch,dinsmore} and adaptive deposition of nanoparticles on responsive microgels~\cite{brasili,brasili2023}. Since the seminal work of Thomson~\cite{thomson}, a core challenge has been to position the interaction that led to a given particle arrangement between its two limits---long-range Coulomb repulsion on one side and short-range excluded volume on the other---and determine whether the surface geometry itself, through its curvature, influences it~\cite{bausch,bowick,jimenez,burke,suncones,guerra,kelleher2017,vutukuri}. Identifying these mechanisms is critical for understanding pattern selection and engineering reconfigurable metamaterials~\cite{prl134,negro2023,carenza,wu2026,moradi,casiulis2026fastgenerationspectrallyshapeddisorder}.

The arrangements of particles on a sphere ($\mathcal{S}^2$) have been characterized through local bond order, topological defects, and the spherical structure factor~\cite{bowick,vitelli2004anomalous,bozic2019,lieu2020topological,vest2018glassy}. The latter measures how strongly a configuration departs from randomness at each angular scale and underlies the classification of hyperuniform states on the sphere~\cite{bozic2019,meyra,bozic2021,brauchart1,brauchart2}. Through potential theory, it has been established that the order in a particle arrangement depends on the underlying interaction: for particles repelling as $r^{-s}$, the limiting arrangement changes at a critical $s$ equal to the dimension $d$ of the surface~\cite{bhs,hardinsaff}---below $s=d$ the particles follow the $s$-equilibrium measure, whereas above it they spread uniformly by area.

The high symmetry of a sphere creates a geometric bottleneck for particles arranged on its surface, as the underlying ordering mechanism leaves no trace in their one-body density~\cite{bhs}. Spreading points uniformly by area (Tammes problem), letting them arrange as charges on a conductor (Thomson problem), or concentrating them where the surface curves most all give the same answer on $\mathcal{S}^2$~\cite{bhs,hardinsaff,landau}. Consequently, as the number of particles approaches infinity, hard spheres and Coulomb charges confined to a spherical surface produce identical mean densities~\cite{bhs,hardinsaff}; distinctions appear only in many-body correlations. The one-body differences are visible to real-space coordinate tracking but not to orientationally averaged scattering. However, if the sphere is stretched into an ellipsoid, these measures become distinct. Hard-core packing spreads particles uniformly by area, curvature-coupled adsorption concentrates them where the surface curves the most, and the conductor, whose surface charge follows a yet different rule, sits between them. The question is whether that distinction can be read from an observable that does not require tracking individual particles.

In this work, we demonstrate that asphericity is not an experimental nuisance to be fitted away, but a powerful diagnostic resource---one that remains underexplored~\cite{wu2026}. Curvature-anisotropic ordering on ellipsoids has been studied in real space~\cite{burke}, where the local structure and defect distribution were shown to respond to the anisotropy of the curvature. What has been missing is a reduction of the competing mechanisms to a single parameter which can be obtained through an orientationally averaged observable. Here, we show that aspherical deformation moves the signature of the ordering mechanism directly into the accessible one-body density, with a single geometric exponent $p$ positioning the arrangement between the conductor and packing limits. We demonstrate how the exponent is realized by two contrasting interaction families---the Riesz $s$-energy, whose limiting measure switches at the marginal power $s=d$, and the screened Coulomb potential, where the characteristic exponent $p$ varies continuously with ionic strength. In the latter case, salt concentration becomes an experimental dial sweeping the particle ordering across the conductor-to-packing crossover while the underlying surface stays fixed. The results of our work are directly applicable to experiments on droplets, vesicles, and microgels, where either the low-$\ell$ structure factor or the radius of gyration can be used to extract the characteristic exponent of the particle arrangement.

\section*{Theory}

\subsection*{Lifting degeneracy with a single exponent}

We parametrize an aspherical shell carrying $N$ particles---a vesicle, droplet, or microgel---as a smooth triaxial ellipsoid $\mathcal{E}$ with semi-axes $a_i$. The ellipsoid coordinates $\mathbf{x}\in\mathcal{E}$ are obtained from unit-sphere coordinates $\mathbf{u}\in\mathcal{S}^2$ by $\mathbf{x} = \bm{M}\mathbf{u}$ with $\bm{M}=\mathrm{diag}(a_i)$. To describe the particle density, we will require two functions, which turn out to be the same function on the ellipsoid. The first is the area distortion of the map $\bm{u}\mapsto\bm{M}\bm{u}$ (the area Jacobian): a patch of solid angle $\dOm_u$ on the unit sphere is mapped onto the ellipsoid with area
\begin{equation}
    \dA = w(\bm{u})\,\Pi_i a_i\,\dOm_u,
\end{equation}
where
\begin{equation}
    w(\bm{u}) = \sqrt{\sum_i u_i^2/a_i^2}.
\end{equation}
The second is the reciprocal support function, defined on the ellipsoid itself:
\begin{equation}
    h(\bm{x}) = (\bm{x}\cdot\hat{\bm{n}})^{-1} = \sqrt{\sum_i x_i^2/a_i^4},
\end{equation}
the inverse of the perpendicular distance from the center to the tangent plane at $\bm{x}$. Substituting $x_i = a_iu_i$ gives $x_i/a_i^2 = u_i/a_i$, hence
$h(\bm{M}\bm{u}) = w(\bm{u})$. The area distortion of the map and the reciprocal support function of the surface are thus the same quantity, one expressed in sphere coordinates and the other in ellipsoid coordinates. Both reduce to $1/R$ on $\mathcal{S}^2$.

An arrangement of $N$ particles on $\mathcal{E}$ gives rise to a particle density $n(\bm{x})$. We consider three measures whose resulting particle density is a distinct power of $h(\bm x)$. Packing of hard-core particles yields $n\propto h^0$ via the poppy-seed bagel theorem~\cite{bhs,hardinsaff}, while classical electrostatics shows that the surface density of charged particles on a conducting ellipsoid scales as $n \propto h^{-1}$~\cite{landau}. The third case is a phenomenological measure for a dilute adsorbed layer whose binding energy depends on the local curvature. This is, for instance, the mechanism by which curvature-sensing proteins localize to the poles of rod-shaped
bacteria~\cite{ramamurthi2009geometric,ramamurthi2009negative,callanjones2011}
and by which defects in cholesteric shells migrate to the poles of an ellipsoid~\cite{negro2023}. For coupling to Gaussian curvature $K_G$, any smooth coupling yields a Boltzmann weight $n\propto\exp(-\mu K_G/\langle K_G\rangle)$, which at modest deformations gives $n\propto h^{4\mu}$ since $K_G = 1/(h^4\Pi_i a_i^2)\propto h^{-4}$~\cite{poelaert2011surface}. Coupling to mean curvature $H=(\sum_ia_i^2-\lVert\bm{x}\rVert^2)/(2h^3\Pi_ia_i^2)$ is less straightforward, but leads to $n\propto h^{-2}$ to leading order in deformation from a sphere~\cite{poelaert2011surface}.

Writing the surface number density as $n(\mathbf{x}) \propto h^{\,p-1}(\mathbf{x})$, the particle density per unit projected solid angle becomes
\begin{equation}
\frac{\dd N}{\dd\Omega_u}=n\frac{\dd A}{\dd\Omega_u}\propto h^{p-1}\, w \propto w^p,
\label{eq:p_family}
\end{equation}
where the exponent $p$ defines the structural regime:
\begin{equation}
p = \begin{cases}
+1 & \text{area-uniform (hard-core packing);} \\
\phantom{+}0 & \text{conductor equilibrium (Thomson);} \\
-3 & \text{curvature-driven localization (model).}
\end{cases}
\label{eq:cases}
\end{equation}
We stress that the last entry is a model choice rather than a derived regime. The curvature weight gives a continuous family $p=4\mu+1$, and our choice of $\mu=-1$, a binding energy of order $k_BT$ per unit of normalized Gaussian curvature (at the weak end of what is measured for curvature-sensing proteins and lipids~\cite{larsen2020,baumgart2011}), is taken only to fix a representative negative value. Unlike the first two measures, it does not arise as the equilibrium of a particle-level energy functional and we do not implement it numerically; its main purpose is to show the negative-$p$ side of the axis, so that the sign of $p$ separates curvature-driven localization from both interaction-driven limits. The exponent $p$ is thus best read as a continuous parameter set by the microscopic coupling, rather than a discrete label.

\begin{figure}[tb]
\includegraphics{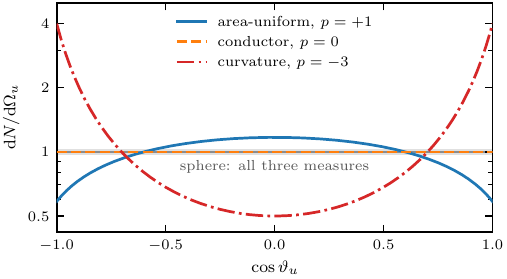}
\caption{The sphere is degenerate; deformation lifts it. Density per unit projected solid angle, $\mathrm{d}N/\dOm_u\propto w^p$, on a spheroid with $c/a=2$ ($\delta=1$) for area-uniform ($p=1$), conductor ($p=0$), and curvature-driven ($p=-3$; model) measures, each normalized to unit solid-angle average. On $\mathcal{S}^2$, all three measures collapse to the grey unity line. Note the logarithmic $y$ scale.}
\label{fig:fig1}
\end{figure}

On a perfect sphere $w$ is constant, so every measure of the form in Eq.~\eqref{eq:cases} reduces to the uniform one and the exponent $p$ leaves no trace in the one-body density (Fig.~\ref{fig:fig1}). Away from sphericity, $p$ quantifies how intensely particles favor flat regions over highly curved tips in the projected density. For example, positive $p$ drives packing toward the equator of a prolate shell, negative $p$ drives localization toward the poles, and $p = 0$ denotes absolute structural indifference. A measurement of $p$ therefore places the arrangement on the axis between conductor and packing order. On its own, the exponent cannot be used to determine the interaction: as we show below, physically distinct interactions can arrive at the same $p$. It is the way $p$ changes as a control parameter is varied that, together with its sign, distinguishes the interactions.

\subsection*{Extracting the exponent from the structure factor}

Analyzing particles on an ellipsoid using the spherical structure factor requires mapping their positions to a reference sphere via a geometric transformation $\bm{T}:\mathcal{E}\to\mathcal{S}^2$. Any such map distorts local area, so a uniform surface density on the ellipsoid becomes inhomogeneous on the sphere. We will mostly work with the affine transformation, which maps coordinates linearly via
\begin{equation}
  \bm{T}_{\rm affine}(\mathbf{x}) = M^{-1}\mathbf{x}
  = \left(\tfrac{x_1}{a_1},\tfrac{x_2}{a_2},\tfrac{x_3}{a_3}\right).
\end{equation}
Alternatively, we will consider a central radial projection
\begin{equation}
  \bm{T}_{\rm radial}(\mathbf{x}) = \mathbf{x}/\|\mathbf{x}\|
  = (x_1,x_2,x_3)\big/\textstyle\sqrt{\sum_i x_i^2}.
\end{equation}
The two send the same surface point to different directions, so the same patch of $\mathcal{E}$ subtends different solid angles under each: we write $\mathrm{d}\Omega_u$ for the solid-angle element of the affine image $\mathbf{u}=\bm{M}^{-1}\mathbf{x}$, and $\mathrm{d}\Omega_r$ for that of the radial image $\mathbf{x}/\|\mathbf{x}\|$. Both integrate to $4\pi$, but they scale differently against the ellipsoid's area element $\mathrm{d}A$.

If we regard a particle configuration mapped from an ellipsoid to the sphere as a density on $\mathcal{S}^2$, $\rho(\Omega)=\sum_k\delta(\Omega-\Omega_k)$, the corresponding spherical structure factor is written as~\cite{bozic2019}
\begin{equation}
    S(\ell) =1+\frac{2}{N}\sum_{j<k}P_\ell(\cos\gamma_{jk}),
    \label{eq:defSl}
\end{equation}
where $P_\ell$ are the Legendre polynomials and $\gamma_{jk}$ is the geodesic distance between particles $j$ and $k$. Taking the ensemble average over configurations in which each particle is drawn from the normalized one-body density $\varrho(\Omega)=N^{-1}\,\mathrm{d}N/\dOm_u$ (with $\int\varrho\,\dOm=1$) splits it into three contributions,
\begin{equation}
\langle S(\ell)\rangle = 1 + \Delta S_\ell + \Gamma_\ell.
\label{eq:avgS}
\end{equation}
The unity is exact for any one-body density; the third term in Eq.~\eqref{eq:avgS}, $\Gamma_\ell$, is the integral over the pair correlation function and is the only one that contains inter-particle correlations. For independent placements, $\Gamma_\ell=0$, whereas $\Gamma_\ell\to-1$ at low $\ell$ for hyperuniform states~\cite{bozic2019}. The middle term of Eq.~\eqref{eq:avgS} is central to this work; we have $\langle P_\ell(\cos\gamma_{jk})\rangle=\iint\varrho(\Omega_1)\varrho(\Omega_2)P_\ell(\hat{\Omega}_1\cdot\hat{\Omega}_2)\dOm_1\dOm_2$, where $\hat{\Omega}$ is a unit vector pointing in the direction $\Omega$. Using the addition theorem for spherical harmonics, we obtain
\begin{equation}
    \Delta S_\ell = (N-1)\,\frac{4\pi}{2\ell+1}\sum_m|\varrho_{\ell m}|^2,
    \label{eq:S_ell}
\end{equation}
with $\varrho_{\ell m}=\int \varrho(\Omega)Y_{\ell m}^*(\Omega)\dOm$. The one-body moments are the ensemble mean of the raw multipole moments, $\varrho_{\ell m}=\langle\rho_{\ell m}\rangle/N$. Importantly, $\Delta S_\ell$ scales linearly with $N$, vanishes for odd $\ell$ by centrosymmetry, and survives even in highly ordered lattices. On a sphere, $\Delta S_\ell$ is invariably zero for $\ell\geqslant1$.

For a uniaxial spheroid with axes $a_1=a_2=a$ and $a_3=c$, the shape parameter is $\delta = c/a - 1$. The projected particle density is axially symmetric, with $\varrho_{\ell0}=\sqrt{(2\ell+1)/4\pi}\int\varrho(\vartheta)P_\ell(\cos\vartheta)\dOm$ and zero for $m\neq0$. Consequently, Eq.~\eqref{eq:S_ell} reduces to $\Delta S_\ell=(N-1)\langle P_\ell \rangle^2$ with $\langle P_\ell\rangle=\int\varrho(\vartheta)P_\ell(\cos\vartheta)\dOm$. Under the affine map, we have $w(\vartheta)=\left(\sin^2\vartheta+\cos^2\vartheta/(1+\delta)^2\right)^{1/2}\approx1-\delta\cos^2\vartheta+\mathcal{O}(\delta^2)$ and so
\begin{equation}
    \varrho(\Omega)=\frac{w^p(\Omega)}{\int w^p(\Omega)\,\dOm}\approx
    \frac{1}{4\pi}\frac{1-p\delta\cos^2\vartheta}{1-p\delta/3}+\mathcal{O}(\delta^2).
\end{equation}
With this, we can write
\begin{equation}
    \varrho(\vartheta)\approx\frac{1}{4\pi}\left[1-\frac{2}{3}\,p\delta\, P_2(\cos\vartheta)\right]+\mathcal{O}(\delta^2)
\end{equation}
Then,
\begin{equation}
    \langle P_2\rangle=\int\varrho P_2\,\dOm = -\frac{2}{15}\,p\,\delta + \mathcal{O}(\delta^2)
    \label{eq:P2}
\end{equation}
and we obtain
\begin{equation}
\Delta S_2 = \frac{4}{225}\,p^2\,(N-1)\,\delta^2+\mathcal{O}(\delta^3).
\label{eq:quadrupole_spheroid}
\end{equation}

Equation~\eqref{eq:quadrupole_spheroid} has a simple interpretation. The quadrupole $\langle P_2\rangle$ measures the excess of particles near the poles over the equator. On a spheroid that excess is proportional to the deformation $\delta$, and the exponent $p$ determines how a given deformation is converted into a given excess---a factor of $-\tfrac{2}{15}\delta p$. The structure factor sees this excess only through its square, because $S(\ell)$ is built from pair distances and is blind to orientation: a pole-heavy and an equator-heavy ellipsoid with the same $|p|$ give the same $\Delta S_2$. The factor $(N-1)$ counts the pairs that share the same one-body bias, which is why the geometric contribution grows with $N$ while the random part of $S(\ell)$ does not. (The radius of gyration, being linear in $\langle P_2\rangle$, provides the sign as well, as we discuss later on  [Eq.~\eqref{eq:Rg}].) Large deformations ($\delta \gtrsim 0.2$) introduce an asymmetry that systematically shifts the quadrupole response, meaning that $p$ must be extracted using the full geometric profile rather than the leading-order contribution of Eq.~\eqref{eq:quadrupole_spheroid}.

\subsection*{From the spheroid to a general ellipsoid}

The result of Eq.~\eqref{eq:quadrupole_spheroid} is generalizable to an arbitrary triaxial ellipsoid with linear axis strains $\varepsilon_i$. Under small triaxial deformations, we can write $a_i=R\,(1+\varepsilon_i)$ (with $\sum \varepsilon_i = 0$ to preserve volume to leading order) where $R$ is fixed and the strains describe shape alone. It follows that
\begin{equation}
  w = \frac{1}{R}\Bigl[1-\sum_i\varepsilon_iu_i^2\Bigr]+\mathcal{O}(\varepsilon^2),
\end{equation}
using $\sum_iu_i^2=1$; hence $w^p\propto1-p\sum_i\varepsilon_iu_i^2$. The quantity $\sum_i\varepsilon_iu_i^2$ is {purely} $\ell=2$: its solid-angle average is $\tfrac13\sum_i\varepsilon_i=0$, so it has no monopole, and being quadratic in $\bu$ it contains nothing above the quadrupole:
\begin{equation}
  \sum_i\varepsilon_iu_i^2
  = \varepsilon_3P_2(\cos\vartheta)+\tfrac12(\varepsilon_1-\varepsilon_2)\sin^2\vartheta\cos2\varphi.
\end{equation}

To first order, $\varrho=\tfrac{1}{4\pi}(1-p\sum_i\varepsilon_iu_i^2)$ and orthogonality gives $\varrho_{2m}=\int\varrho\, Y_{2m}^*\dOm=-\tfrac{p}{4\pi}\int\sum_i\varepsilon_iu_i^2\,Y_{2m}^*\dOm$. It follows that
\begin{equation}
  \varrho_{20}=-\frac{p\,\varepsilon_3}{2\sqrt{5\pi}},\quad
  \varrho_{2,\pm2}=-\frac{p\,(\varepsilon_1-\varepsilon_2)}{4}\sqrt{\frac{2}{15\pi}},
  \label{eq:denm}
\end{equation}
and the power of the $\ell=2$ component collapses to a rotational invariant:
\begin{equation}
  \sum_m\lvert\varrho_{2m}\rvert^2=\frac{p^2}{30\pi}\sum_i\varepsilon_i^2=\frac{5}{64\pi}p^2\kappa^2 .
  \label{eq:rlm}
\end{equation}
We introduced $\kappa^2=\tfrac32\mathrm{tr}(\hat{\bm\Lambda}^2)/(\mathrm{tr}\bm\Lambda)^2$ as the relative shape anisotropy of the \emph{surface} gyration tensor $\Lambda_{ij}=A^{-1}\!\oint x_ix_j\,\dA$, with $\hat{\bm\Lambda}$ the deviatoric part. We use the surface tensor throughout because the particles live on the surface; the more common solid gyration tensor gives a $\kappa^2$ larger by $25/16$ at small deformation (Appendix~\ref{app:B}). Substituting Eq.~\eqref{eq:rlm} into Eq.~\eqref{eq:S_ell} for $\ell=2$ then yields 
\begin{equation}
\Delta S_2 = \frac{1}{16}\,p^2\,(N-1)\,\kappa^2+\mathcal{O}(\varepsilon^3).
\label{eq:central_relation}
\end{equation}

\begin{figure}[tb]
\includegraphics{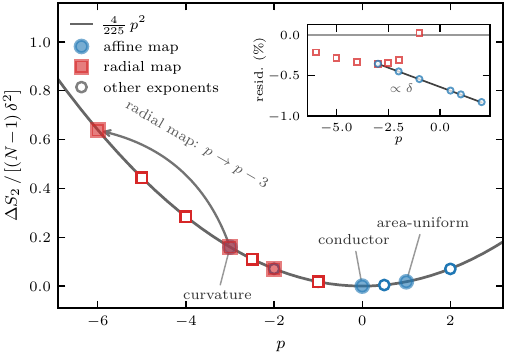}
\caption{Normalized $\Delta S_2$ of every reference measure, under both coordinate projections, falls on a single parabola in $p$ (line). Evaluated on a spheroid with $\delta=0.005$. Filled symbols are the three reference measures of Eq.~\eqref{eq:cases}; open symbols represent arbitrary exponents that demonstrate the validity of the parabola between the physical values. Circles use the affine map ($p$), squares the radial one ($p\to p-3$); the arrow shows that shift for the model curvature measure. Inset: residual deviation from Eq.~\eqref{eq:quadrupole_spheroid}, scaling as $\mathcal{O}(\delta)$. The deviation of the affine map can be predicted to scale as $-3\delta(\tfrac37+\tfrac{4p}{63})$ (line); deviation of the radial map departs from it.}
\label{fig:fig2}
\end{figure}

The exponent that enters Eqs.~\eqref{eq:quadrupole_spheroid} and \eqref{eq:central_relation} is that of the particle density \emph{as projected onto the reference sphere}, and it depends on the map while the physical exponent $p$ of Eq.~\eqref{eq:p_family} does not. Under the affine map $\bm{T}_\mathrm{affine}$ the projected exponent is $p$ itself; under the radial map $\bm{T}_\mathrm{radial}$ the geometric spreading factor $\|\bm x\|^3$ shifts it to $p-3$. Note that $rh=1+\mathcal{O}(\delta^2)$, so both $r^3\propto w^{-3}$ and the resulting shift of the exponent by three---the dimension of the space the surface is embedded in---hold to that order. Numerical evaluations across different reference measures validate that the quadratic form of Eq.~\eqref{eq:quadrupole_spheroid} is universal in $p$ to leading order in deformation regardless of the mapping (Fig.~\ref{fig:fig2}). Throughout this work, $p$ denotes the physical exponent and the shift is written out explicitly where the radial map is used. 

\section*{Results}

\subsection*{Physical realizations}

To test the continuum geometric predictions of the theory in discrete systems, we study the configurations of particles on an aspherical surface that interact via two contrasting interaction families: \emph{(i)} the generalized Riesz potential $V_\mathrm{R} = r^{-s}$ characterized by the exponent $s$, and \emph{(ii)} the screened Coulomb (Debye--H\"uckel) potential $V_\mathrm{DH} = \frac{1}{r}\, e^{-r/\lambda_D}$, with $\lambda_D$ the Debye screening length. In both cases, $r$ denotes the three-dimensional chord distance---a physical choice, since potentials acting through space behave differently from those acting along geodesics~\cite{law}. The configurations are obtained by minimizing the respective energies by projected gradient descent at zero temperature (Appendix~\ref{app:F}); thermal motion would bias particles toward the area-uniform measure, so a finite-temperature $p$ should interpolate between the value reported here and $p=1$.

\begin{figure*}[t]
\includegraphics{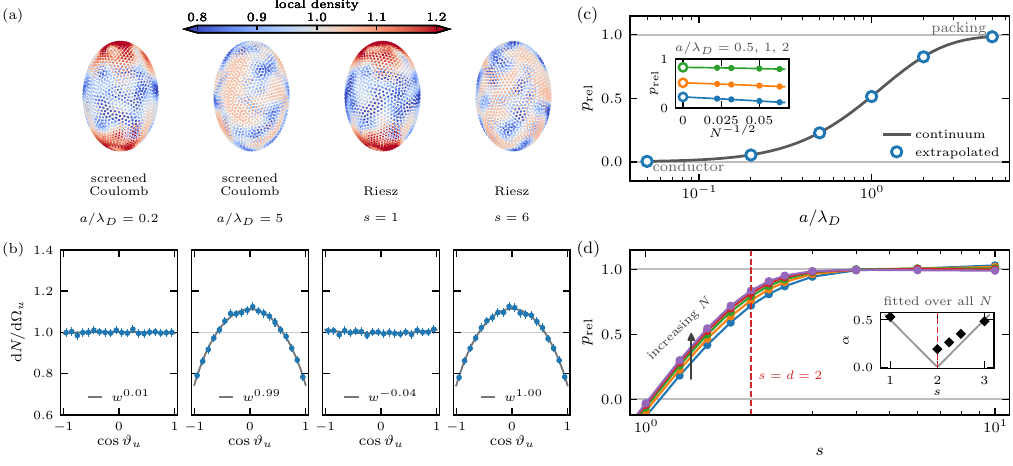}
\caption{The conductor-packing crossover in two interaction families. Throughout, we use a spheroid with $c/a=1.5$ ($\delta=0.5$), analyzed with an affine map. (a) Representative configurations of $N=2000$ particles interacting via a screened Coulomb or Riesz potential. The parameters $a/\lambda_D$ and $s$ sit at the two ends of the conductor-to-packing crossover. Particles are colored by the local areal density, estimated from the inverse square of the distance to the sixth-nearest neighbor and averaged over each particle's neighborhood. (b) Corresponding particle density per unit projected solid angle, pooled over $24$ independent configurations (symbols), against the profile $w^{\,p}$ (lines) whose exponent is fixed by the quadrupole of the pooled samples. (c) Exponent $p_\mathrm{rel}$ for screened Coulomb configurations as a function of screening $a/\lambda_D$. Symbols are $N\to\infty$ limits extrapolated from finite $N$ ($12$ realizations each at $N=250$, $400$, $1000$, and $2000$); standard errors are smaller than the symbols. The line shows the continuum equilibrium measure. Inset: the extrapolation of $p_{\rm rel}$ against $N^{-1/2}$ at three couplings, with a two-term fit (lines) and the $N\to\infty$ limits (open circles). (d) Exponent $p_\mathrm{rel}$ as a function of the exponent $s$ of the Riesz $s$-energy family for $12$ configurations each at $N=250$, $500$, $1000$, $2000$, and $4000$ (arrow). The limiting measure changes character at $s=d=2$ (dashed line). Inset: convergence exponent $\alpha$, fitted across the shown $N$, as a function of $s$ and compared against $\lvert1-s/d\,\rvert$ (line).}
\label{fig:fig3}
\end{figure*}

For simplicity, we limit ourselves to particles confined to a spheroid. Because the minimized configurations are hyperuniform~\cite{bozic2019}, $\Gamma_\ell\neq0$ and we therefore work directly with $\langle P_2\rangle$ on a scale fixed by two reference measures and define the relative exponent as $p_\mathrm{rel}=\langle P_2\rangle/\langle P_2\rangle_{\rm area}$ (using $\langle P_2\rangle_{\rm conductor}=0$). Thus $p_\mathrm{rel}=0$ is the conductor measure and $p_\mathrm{rel}=1$ the area-uniform one; $p_\mathrm{rel}$ coincides with $p$ only as $\delta\to0$. Figure~\ref{fig:fig3}a shows how the Riesz and the screened Coulomb particle configurations look at the two ends of $p_\mathrm{rel}$, and Fig.~\ref{fig:fig3}b confirms that they are indeed members of the $w^{\,p}$ family---the measured densities follow the predicted angular distribution closely.

Figure~\ref{fig:fig3}c shows the coupling exponent $p_\mathrm{rel}$ of minimal-energy configurations of particles interacting with a screened Coulomb potential. The exponent is extrapolated to $N\to\infty$ from finite-$N$ configurations by fitting $p_\mathrm{rel}(N)=p_\mathrm{rel}^{\infty}-AN^{-1/2}-BN^{-1}$, in which $A$ and $B$ are fit coefficients and the leading rate is $\alpha=\lvert1-s/d\,\rvert=1/2$ as expected in the Coulomb limit (where $d=2$ is the dimension of the spheroid)~\cite{bhs}. The exponent crosses over smoothly from the conductor value to the area-uniform value as the screening $a/\lambda_D$ increases. What is more, the extrapolated exponent agrees with the continuum equilibrium measure, obtained from the constant-potential condition
\begin{equation}
    \int e^{-\lvert\bx-\by\rvert/\lambda_D}\lvert\bx-\by\rvert^{-1}n(\by)\,\dA(\by)=\text{const}
\end{equation}
on the same spheroid (full line in Fig.~\ref{fig:fig3}c). Ionic strength is thus a tuning parameter that continuously guides the particle order on a spheroid from Thomson-like (low screening) to Tammes-like (high screening).

Configurations of particles interacting via the Riesz potential depend on the exponent $s$ that similarly interpolates between long-range (Coulomb) electrostatics ($s=1$) and local close-packing ($s \to \infty$). The configurations are expected to follow the $s$-equilibrium measure for $s<d$ and the surface (area-uniform) measure for $s\geqslant d$~\cite{bhs,hardinsaff}, with a transition when $s=d=2$. Our minimal-energy configurations confirm that $p_{\text{rel}}$ crosses over continuously from $p_\mathrm{rel}=0$ at $s=1$ to approach $p_\mathrm{rel}=1$ for $s \geqslant 2$ as $N\to\infty$ (Fig.~\ref{fig:fig3}d). The convergence rate of the empirical measure scales as $N^{-\alpha}$, where the exponent $\alpha$ hits a minimum precisely at the marginal case $s=d=2$. Convergence therefore slows down markedly at the marginal point, consistent with the logarithmic bottleneck that the energy asymptotics predict~\cite{bhs}; the vanishing $\alpha=\lvert1-s/d\,\rvert$ they predict is not reached at accessible $N$. In contrast, the screened Coulomb potential never enters the $s\geqslant d$ regime, and no sharp transition occurs---the effective coupling exponent $p_\mathrm{rel}$ varies smoothly as a continuous function of ionic strength (Fig.~\ref{fig:fig3}c).

The two interaction families demonstrate what can be gleaned from a single value of $p_{\rm rel}$. The particle configurations in both span the same interval of values, reaching $p_{\rm rel}\approx1$ in the appropriate limits (Fig.~\ref{fig:fig3}c,d). As a consequence, screened Coulomb repulsion at high screening ($a/\lambda_D\gtrsim3$) and a Riesz-type interaction with $s\geqslant d$ will lead to a similar value of $p_{\rm rel}$. They can be separated by the response to the variation in $\lambda_D$ and $s$, respectively. The exponent $p_\mathrm{rel}$ of Riesz configurations switches at the marginal power, and its convergence rate collapses there ($\alpha\approx0.1$ at $s=d$). The screened Coulomb exponent $p_\mathrm{rel}$ drifts continuously with $a/\lambda_D$, and its rate, measured against the continuum measure in the same way, never slows: $\alpha=0.53$--$0.54$ across the conductor side of the crossover and rising to $\alpha\approx0.9$ under strong screening, where the interaction becomes short-ranged on the scale of the spheroid. An experiment that scans a control parameter, rather than measuring at a single setting, could therefore recover the interaction family from the trajectory of $p$.

\subsection*{Experimental detection threshold}

The crossover seen in Fig.~\ref{fig:fig3}c is driven by a quantity that can be experimentally controlled. For charge-stabilized colloids confined to stretched vesicles or liquid crystal droplets~\cite{dinsmore} or for nanoparticles assembling on polystyrene beads under a screened Coulomb interaction~\cite{moradi}, the inverse screening length $\lambda_D^{-1}$ is set by the salt concentration $c$, $\lambda_D^{-1}\propto\sqrt c$. Tuning the ionic strength thus provides an experimental dial to continuously translate $p$ across the conductor-to-packing spectrum. If the surface shape, characterized by $\kappa^2$, remains unchanging, any shift in the low-$\ell$ signal tracks the reorganization of the particle layer across the conductor-to-packing crossover.

How large a deformation such an experiment requires is set by the noise due to the  limited number of realizations of the particle configurations and their finite size. For independent, uniform placements of particles, each multipole coefficient $\rho_{\ell m}$ is a sum of $N$ independent terms of zero mean and variance $\langle\lvert Y_{\ell m}\rvert^2\rangle=1/4\pi$. By the central limit theorem it is Gaussian with variance $\frac{N}{4\pi}$, and the $2\ell+1$ modes are uncorrelated due to the orthogonality of the spherical harmonics. Consequently, $\frac{4\pi}{N}\sum_m\lvert\rho_{\ell m}\rvert^2$ is a sum of $\nu=2\ell+1$ squared unit Gaussians---a $\chi_\nu^2$ variable---and $S(\ell)=\chi_\nu^2/\nu$. Since $\chi_\nu^2$ has a mean of $\nu$ and variance $2\nu$, we have $\langle S(\ell)\rangle=1$ and $\mathrm{Var}[S(\ell)]=2/(2\ell+1)$.

When independent placements are made from an anisotropic $\varrho$, $\langle\rho_{\ell m}\rangle=N\varrho_{\ell m}\neq0$, and the sum of squares becomes a non-central $\chi^2$ with noncentrality $\eta=\frac{4\pi}{N}\sum_m\lvert N\varrho_{\ell m}\rvert^2$. The mean and variance become $\nu+\eta$ and $2\nu+4\eta$, respectively, and thus $\langle S(\ell)\rangle=1+\eta/\nu=1+\Delta S_\ell$ and
\begin{equation}
  \mathrm{Var}[S(\ell)] =
  \frac{2}{2\ell+1}\bigl(1 + 2\,\Delta S_\ell\bigr).
  \label{eq:varS}
\end{equation}
For correlated (hyperuniform) arrangements, $\lim_{\ell\to0}S(\ell)=0$ by definition~\cite{bozic2019} and the low-$\ell$ modes of the particle density fluctuate far less than the modes of independent placements, $\mathrm{Var}[\rho_{\ell m}]\to0$ relative to $N$. In this respect, Eq.~\eqref{eq:varS} is the variance of a particle layer with the same one-body density and no correlations and thus represents a conservative bound; any hyperuniform correlations can only lower it.

When $S(2)$ is averaged over $K$ independent realizations (particle arrangements), the standard deviation $\sigma$ is obtained as $\sigma^2=\mathrm{Var}[S(2)]/K=\frac{2}{5K}(1+2\Delta S_2)$. If we demand that the signal be $z$ standard deviations above zero, $\Delta S_2=z\sigma$ and
\begin{equation}
  \Delta S_2^{\,2} = \xi\,(1 + 2\Delta S_2),
  \qquad \xi = \frac{2z^2}{5K}.
  \label{eq:quad}
\end{equation}
Its positive root, $\Delta S_2 = \xi + \sqrt{\xi(\xi+1)}$, equated to Eq.~(\ref{eq:central_relation}), gives 
\begin{equation}
\kappa^2_{\text{min}} = \frac{16\left[\,\xi+\sqrt{\xi(\xi+1)}\,\right]}{p^2(N-1)},
\quad \xi=\frac{2z^2}{5K}.
\label{eq:threshold}
\end{equation}
Under the radial map, common in experimental setups, one needs to replace $p\to p-3$. The $\kappa^2_{\min}$ presupposes no shape family---for a spheroid, we have $|\delta|_{\min} = \frac{15}{8}\sqrt{\kappa^2_{\min}}$. The threshold of Eq.~\eqref{eq:threshold} is thus the deformation at which a \emph{disordered} layer of the same mean density becomes detectable; an ordered layer clears it at far smaller deformation, and its residual scatter reflects the sample-to-sample variation of the exponent itself rather than pair noise.

As a worked example, we consider a recent study of nanoparticles adsorbed on thermoresponsive microgels~\cite{brasili,brasili2023}. There, anionic gold particles (diameter $20$~nm, charge $-35e$, Debye length $\sim20$~nm) are adsorbed to cationic pNIPAM microgels with $\sim10^2$ particles adsorbed per microgel. The microgel undergoes a temperature-dependent collapse, and its hydrodynamic radius shrinks from $R_\mathrm{H}\approx286$~nm to $\approx110$~nm; importantly, during the collapse the corona shrinks so the nanoparticle layer goes from a thick shell to a genuinely two-dimensional one. Molecular dynamics simulations allow the determination of the spherical structure factor by projecting each adsorbed nanoparticle radially (i.e., using $\bm{T}_{\rm radial}$) onto the unit sphere and averaging over five independent microgels ($K=5$). The resulting low-$\ell$ plateau goes from $S\approx0.45$ in the swollen phase to $S\approx0.1$ in the collapsed phase (Fig.~3C in Ref.~\cite{brasili}). Both values sit above the hyperuniform limit $\lim_{\ell\to0}S(\ell)=0$, closer to a liquid and consistent with the authors' Voronoi analysis~\cite{brasili}. The predicted exponent of an area-uniform layer ($p=1$) under the radial map is $p-3=-2$, and Eq.~\eqref{eq:central_relation} gives $\Delta S_2\approx25\kappa^2$ for $N=100$, with $\kappa^2$ the shape anisotropy of the microgel. To explain the entire high-temperature plateau, Eq.~\eqref{eq:central_relation} yields for $\Delta S_2=0.1$ a shape anisotropy of $\kappa^2=0.004$; equivalently, for a spheroid, $\delta\approx0.12$, a $12\%$ elongation. A simulated polymer network of $10^4$ monomers is unlikely to be spherical to within a few percent, so the geometric contribution is plausibly not negligible against the $0.1$ plateau; the question is how much of the plateau is explained by the asphericity. Furthermore, computing $S(\ell)$ at different $N$ would separate the two contributions, since $\Delta S_\ell\propto N-1$ at fixed shape while $\Gamma_\ell$ does not scale with $N$.

\subsection*{Obtaining the exponent from the radius of gyration}

Determining $S(\ell)$ requires particle coordinates, which are easy to obtain in simulations but more difficult in experiment. Experimentally, however, analogous information needed to extract $p$ can be obtained from an orientationally averaged scattering intensity $I(q)$ of a randomly oriented suspension of shells (spheroids/ellipsoids in our case), where $q$ is the magnitude of the scattering vector. Expanding the plane wave in spherical waves and averaging over the scattering vector directions $\hat{\mathbf{q}}$ yields~\cite{pedersen1997analysis}:
\begin{equation}
  I(q) = 1 + 4\pi(N - 1)\sum_{\ell m} |B_{\ell m}(q)|^2, \label{eq:D2}
\end{equation}
where $B_{\ell m}(q) = \int \varrho_r j_\ell(qr) Y^*_{\ell m} \mathrm{d}\Omega_r$. Here, $\varrho_r$ represents the one-body particle density per unit radial solid angle, making the central radial projection the natural mapping scheme for scattering. On a sphere, $r$ is constant and Eq.~\eqref{eq:D2} reduces to $I(q)=\sum_\ell(2\ell+1)j_\ell^2(qR)S(\ell)$, recovering the standard relation $I(0)=N$. On an ellipsoid the radial weight $j_\ell(qr)$ varies with direction and no such factorization occurs. Because the radial density scales as $\varrho_r \propto h^p r^3$, separating out the purely geometric surface volume factor yields $h^p r^3 = h^{p-3}(rh)^3 = w^{p-3}$ to leading order in deformation, $rh = 1 + \mathcal{O}(\delta^2)$. This geometric area distortion shifts the exponent to $p\to p - 3$, meaning area-uniform packing ($p = 1$)  projects to $\varrho_r \propto w^{-2}$, $\langle P_2 \rangle > 0$ on the reference sphere. 

In the small-$q$ limit, Eq.~\eqref{eq:D2} simplifies to $I(q)/I(0) = 1 - q^2R^2_g/3 + \mathcal{O}(q^4)$. The Guinier radius is measured about the centroid of the particles $\bm{\overline{x}}$~\cite{feigin1987structure}, $R_g^2=\langle r^2\rangle-\lvert\bm{\overline{x}}\rvert^2$, whereas $\langle r^2\rangle=\int\varrho_rr^2\dOm_r$ is taken about the center of the underlying shell. We have $\bm{\overline{x}}=\frac{1}{N}\sum_k\bm{x}_k$ and $\lvert \bm{x}_k\rvert\approx R$ for small deformations. Hence, for $N$ particles, $\lvert\bm{\overline{x}}\rvert^2\approx\frac{R^2}{N}S(1)$ and $R_g^2=\langle r^2\rangle\big(1-S(1)/N\big)$. For independent particle placements, $S(1)=1$ and the correction is $1/N$; for an ordered layer, $S(1)\to0$ and it vanishes. Since the ellipsoid is symmetric under reflection in its principal planes, and the particle density $\varrho_r$ shares this symmetry, the odd-$\ell$ components of $r^2(\Omega_r)$ vanish together with the $m=\pm1$ and $\sin 2\varphi$ components at $\ell=2$. Measuring $\vartheta$ from the $a_3$ axis and $\varphi$ from the $a_1$ axis, we expand the shape of the ellipsoid as
\begin{equation}
r^2(\Omega_r) = u_0 + u_2 P_2(\cos\vartheta)
  + u_{22}\sin^2\vartheta\cos 2\varphi + \ldots,
\label{eq:r2expansion}
\end{equation}
with  $u_2 = \frac{5}{4\pi}\oint r^2 P_2\,\mathrm{d}\Omega_r$, $u_{22} = \frac{15}{16\pi}\oint r^2 \sin^2\vartheta\cos 2\varphi\,\mathrm{d}\Omega_r$, and $u_0$ the solid-angle average of $r^2$. Using $\langle r^2\rangle=\int\varrho_r r^2\,\mathrm{d}\Omega_r$, we obtain
\begin{equation}
R_g^2 = \left(1-\frac{S(1)}{N}\right)
  \Big[u_0 + u_2\langle P_2\rangle
  + u_{22}\langle\sin^2\vartheta\cos 2\varphi\rangle\Big]
  + \mathcal{O}(\varepsilon^4).
\label{eq:Rg}
\end{equation}
For a spheroid, $u_{22}=0$ and only the first anisotropic term remains.

To leading order in $\varepsilon_i$, $r^2 = R^2[1 + 2\varepsilon_3 P_2(\cos\vartheta) + (\varepsilon_1-\varepsilon_2)\sin^2\vartheta\cos 2\varphi]
+ \mathcal{O}(\varepsilon^2)$, so that $u_2 = 2R^2\varepsilon_3$ and $u_{22} = R^2(\varepsilon_1-\varepsilon_2)$. The density moments follow from Eq.~\eqref{eq:denm} with $p\to p-3$: $\langle P_2\rangle = -(p-3)\varepsilon_3/5$ and $\langle\sin^2\vartheta\cos 2\varphi\rangle = -\tfrac{2}{15}(p-3)(\varepsilon_1-\varepsilon_2)$. The two anisotropic terms combine into $-\tfrac{4}{15}(p-3)R^2\sum_i\varepsilon_i^2$, the same rotational invariant as in Eq.~\eqref{eq:rlm}, and thus
\begin{equation}
R_g^2 = \left(1-\frac{S(1)}{N}\right)
  \left[u_0 + \frac{5}{8}\,(3-p)\,R^2\kappa^2\right]
  + \mathcal{O}(\varepsilon^3).
\label{eq:Rgkappa}
\end{equation}
Equation~(\ref{eq:Rgkappa}) parallels Eq.~\eqref{eq:central_relation}: the shape again enters only through $\kappa^2$, but now linearly in the projected exponent $p-3$ rather than through its square. On a perfect sphere ($\kappa^2=0$), $R_g$ becomes entirely independent of $p$, hiding the degenerate ordering mechanism; on an
ellipsoid, the anisotropic terms of Eq.~(\ref{eq:Rg}) remove this degeneracy. The contribution $u_0 = R^2 + \mathcal{O}(\varepsilon^2)$ does not depend on $p$, but it is of the same order as the signal and has to be fixed from the bare shell geometry.

As an example, take $\delta=0.2$, where $u_2/u_0\approx0.25$. Under the radial map, area-uniform packing ($p=1$, projected exponent $p-3=-2$) gives $u_2\langle P_2\rangle/u_0\approx1.2\%$ and the conductor measure ($p=0$, projected exponent $-3$) $\approx1.8\%$. Distinguishing between the area-uniform and conductor states via this method requires resolving a narrow $0.6\%$ differential in $R_{g}^{2}$ for a fairly moderate deformation. Consequently, utilizing $R_{g}$ as a practical probe demands rigorous suppression of experimental polydispersity and precise independent characterization of the bare shell geometry (coefficients $u_{\ell }$). Additionally, the finite-$N$ factor $S(1)/N$ has to be fixed from the measured $S(1)$, or $N$ must be large enough ($N\gtrsim10^3$) so that it drops below $0.1\%$. In return, $R_g^2$ is linear in $\langle P_2\rangle$ [Eq.~\eqref{eq:Rg}] and therefore retains the sign of $p$ that $\Delta S_2$ discards. Experiments performed at different screening strengths could ease the accuracy demands, as any drift in $R_g^2$ as a function of ionic strength would track the reorganization of the particles---and thus $p$---alone.

\section*{Discussion and conclusion}

We have shown that aspherical deformation, usually a correction to fit away, is what makes the ordering mechanism of a particle arrangement visible: measures that a sphere renders indistinguishable are separated by a single characteristic exponent $p$ on an ellipsoid. The value of the exponent is recoverable from the quadrupole of the spherical structure factor [Eq.~\eqref{eq:central_relation}] and from the radius of gyration [Eq.~\eqref{eq:Rg}]. The latter also provides access to $\langle P_2 \rangle$, thus allowing the sign of $p$ to be identified alongside its value. This is relevant since the curvature-weighted measure [Eq.~\eqref{eq:cases}] distinguishes itself from both interaction-driven (conductor and packing) limits by sign alone. Another way to resolve the sign of $p$ is to actively break the orientational symmetry; experimentally, this can be achieved by applying external uniform magnetic fields to liquid crystal droplets~\cite{Ettinger2023} or utilizing microfluidic shear flows to orient lipid vesicles~\cite{McWhirter2009}.

Two interaction families show how particle configurations realize the exponent in complementary ways: the Riesz $s$-energy, whose limiting measure switches to the area-uniform value at $s=d=2$, and the screened Coulomb potential, where the characteristic exponent varies continuously with the salt concentration. The numerical results presented in Fig.~\ref{fig:fig3} were obtained for spheroids for computational convenience. Nonetheless, Eq.~\eqref{eq:central_relation} is derived for a general triaxial ellipsoid and written in $\kappa^2$, so it is not tied to any particular shape family. The $\ell=2$ power is a rotational invariant of the axis strains, which is why $\kappa^2$ suffices; at the same time, prolate, oblate and triaxial shells of equal $\kappa^2$ remain indistinguishable in $S(2)$ once orientation is averaged away.

The characteristic exponent $p$ that is the center of our framework can be utilized in experiments in different ways. For charge-stabilized colloids at fluid interfaces~\cite{kelleher2017,guerra}, where ionic strength already tunes the phase behavior on flat and spherical droplets, it predicts that on an aspherical droplet the same control reorganizes the one-body density at fixed shape, and Eq.~\eqref{eq:threshold} gives a conservative estimate of how aspherical the droplet must be for that reorganization to register in an orientationally averaged measurement. For nanoparticle-decorated microgels~\cite{brasili2023,brasili}, it runs the other way: it converts the shape of the carrier into a definite proportion of the low-$\ell$ structure factor, so that a measured plateau can be partitioned between the particle layer, the carrier shape, and the colloidal state of the complexes~\cite{capocefalo2026} rather than attributed to the first by default. And because the one-body density is itself a source of topological defects~\cite{yao2013,burke}, $p$ also fixes the defect-generating field that an aspherical substrate imposes independently of its Gaussian curvature. In each case the quantity to be measured is the same, $\Delta S_2$ or $R_g^2$, and the shape enters through the same invariant, $\kappa^2$.

The present approach can be positioned relative to the defect-based characterization of curved crystals~\cite{bausch,bowick,burke,jimenez,suncones}. The decomposition of Eq.~\eqref{eq:avgS} separates the two approaches: scars, dislocation clouds and coordination defects are many-body objects and enter $\Gamma_\ell$, the term that survives on a sphere, whereas the exponent $p$ enters $\Delta S_\ell$, the term that is identically zero on a sphere. The two are not independent: \citet{yao2013} showed that a density gradient by itself creates defects as if it were an additional curvature. This coupling vanishes for area-uniform packing ($p=1$), where defects follow the Gaussian curvature alone~\cite{burke}, and grows with $|p-1|$ away from it~\cite{yao2013}, so that a conductor layer should show defects displaced from where curvature by itself would place them. And while defects could in principle serve as a probe of $p$ through that displacement, their number grows only as $\sqrt N$, their response is filtered through the elastic constants of the layer, and on an aspherical surface their identification depends on the metric of the tessellation. By contrast, $\Delta S_2$ grows linearly with $N$, relates to $p$ through Eq.~\eqref{eq:central_relation} alone, and is a scalar of the one-body density. The two diagnostics are complementary rather than competing: defects resolve the local order and its coupling to curvature, the structure factor resolves the measure that order is built on.

Several conditions bound the validity of this framework. First, the surface must be a smooth convex quadric; for more complex shapes, the area Jacobian $w$ and the support function $h$ decouple, breaking the single-exponent reduction. Second, the particles are treated as structureless points on a rigid substrate: finite-size core corrections enter only via modifications to the mean density profile, and particles that deform the surface they sit on~\cite{vahid2017} fall outside the family altogether. Furthermore, the exponents of Fig.~\ref{fig:fig3} are zero-temperature values. For charge-stabilized colloids on micron-scale droplets the correction is expected to be small, but for weakly charged particles on small shells the layer will lie closer to area-uniform than the interaction alone would place it. Next, the scattering derivation for the radius of gyration rests on $r^2$ being dominated by its $\ell\leqslant2$ content, which holds for modest deformations. Lastly, the $(2\ell+1)$ degeneracy splits linearly in $\ell$ under deformation, so a hyperuniformity classification built on the structural peak $\ell_0=\pi\sqrt{N/3}$~\cite{bozic2019,bozic2021} becomes ill-defined on an aspherical shell; the $\ell=2$ diagnostics used here are unaffected. This decomposition also bears on spectral design: optimization methods that impose target spectra on point patterns currently presuppose a perfect sphere~\cite{casiulis2026fastgenerationspectrallyshapeddisorder}, and Eq.~\eqref{eq:central_relation} quantifies the geometric contribution that any real, slightly aspherical substrate injects at low $\ell$.

These conditions mark the edges of the framework's validity; within them, the exponent is indicative of the interaction underlying a particle arrangement, exact for the limiting measures and tracing the crossover between them, and it is read from observables that experiments already measure. The most direct test of the framework is the one it was built for: a charge-stabilized monolayer on an elongated droplet or vesicle, scanned in ionic strength at fixed shape, with $R_g$ or aligned $S(2)$ determined at each step. A trajectory that runs from the conductor value toward the packing value, at the rate the continuum measure predicts, would confirm that the one-body density can be steered by salt alone.

\section*{Conflicts of interest}

There are no conflicts to declare.

\section*{Data availability}

Figures 1 and 2 can be reproduced directly from the presented equations. The data reproducing Figure 3 will be made publicly available on Zenodo upon publication of the manuscript.

\acknowledgments

I thank Simon Čopar for valuable comments. The work was supported by the Slovenian Research and Innovation Agency (research project No.\ J1-60002 and research core funding No.\ P1-0055).

\bibliography{references}

\onecolumngrid 
\section*{APPENDIX}
\vspace{0.2cm}
\twocolumngrid

\appendix

\setcounter{equation}{0}
\renewcommand{\theequation}{A\arabic{equation}}
\section{The two gyration tensors}
\label{app:B}

The shape anisotropy of a general ellipsoid $\kappa^2$ depends on whether one uses the surface or the solid gyration tensor to define it. Both are the second moment of the mass distribution about the centroid, $\Lambda_{ij} = \langle x_ix_j\rangle$, differing only in what carries the mass. The solid gyration tensor is $\Lambda^{\rm sol}_{ij} = \frac{1}{V}\int_{\mathcal{E}} x_ix_j\,\mathrm{d}V = \frac{a_i^2}{5}\,\delta_{ij}$. The surface gyration tensor is $\Lambda^{\rm surf}_{ij} = \frac{1}{A}\oint_{\partial\mathcal{E}} x_ix_j\,\dA$; it involves elliptic integrals in general and is evaluated by quadrature here. From either, the {relative shape anisotropy} is the standard dimensionless combination $\kappa^2 = \frac{3}{2}\text{tr}(\hat{\mathbf{\Lambda}}^2)/(\text{tr}\mathbf{\Lambda})^2$ with $\hat{\bm{\Lambda}} = {\bm{\Lambda}} - \tfrac13\bigl(\operatorname{tr}{\bm{\Lambda}}\bigr)\mathbbm{1}$, i.e.\ the normalized magnitude of the deviatoric part. It vanishes for a sphere; for a spheroid, $\kappa^2_{\rm surf} = \tfrac{64}{225}\delta^2 + \mathcal{O}(\delta^3)$ and $\kappa^2_{\rm sol} = \tfrac{4}{9}\delta^2 + \mathcal{O}(\delta^3)$, and thus $\kappa^2_{\rm sol}/\kappa^2_{\rm surf} \to \tfrac{25}{16}$. Consequently, Eq.~\eqref{eq:central_relation} reads
\begin{equation}
  \Delta S_2
  = \frac{p^2(N-1)\,\kappa^2_{\rm surf}}{16}
  = \frac{p^2(N-1)\,\kappa^2_{\rm sol}}{25}.
\label{eq:ratio_conv}
\end{equation}
Equation~\eqref{eq:ratio_conv} holds for small $\delta$; the ratio $\kappa^2_{\rm sol}/\kappa^2_{\rm surf}$ decreases at larger deformation. 

\setcounter{equation}{0}
\renewcommand{\theequation}{B\arabic{equation}}
\section{Numerical minimization}
\label{app:F}

Configurations minimize the stated energy by projected gradient descent with a geometrically decaying step, using the three-dimensional chord distance. The quantity to converge in is the total path length a particle may travel, not the number of steps. With $\text{step}_k=\text{step}_0\,\lambda^k$ the path length is bounded by $\text{step}_0/(1-\lambda)$ however many steps are taken---so running longer at fixed
$\lambda$ tests only whether that bound has been approached. The default we use, $\text{step}_0=0.02\,\min_i a_i$ and $\lambda=0.999$ over $3000$ steps, permits a path of $19\min_i a_i$; doubling it (via $\lambda=0.9995$ and $6000$ steps) moves the fitted exponent by $3\times10^{-4}$ at $N=1000$ and $1\times10^{-4}$ at $N=4000$. Doubling the step count alone would have raised the bound by $5\%$. Further controls on the deformation, the calibration, and a path-length check at the largest size all pass, and adequacy is verified separately for each observable, since a path length sufficient for a mean-density quantity need not be sufficient for one measuring local order. For the particles interacting via Riesz potential, across twelve independent initial conditions at fixed $(N,s)$ the fitted $p_{\rm rel}$ varies by $\lesssim0.02$ over $1\leqslant s\leqslant3$, below the finite-size spread in Fig.~\ref{fig:fig3}d; the one-body density is far less basin-sensitive than the energy itself, which is what makes the exponent a robust observable. For the screened Coulomb interaction, the leading $N^{-1/2}$ term was checked by measuring the rate directly, with the same definition used for the Riesz family: the decay of $|p_{\rm rel}^{\rm cont}-p_{\rm rel}(N)|$ with $N$,
where $p_{\rm rel}^{\rm cont}$ is the continuum equilibrium measure of Fig.~\ref{fig:fig3}c, bootstrapped over the twelve seeds gives $\alpha=0.53$, $0.53$, $0.53$, $0.54$, $0.56$ and $0.93$ for $a/\lambda_D=0.05$–$5$, the values quoted in the main text.

\end{document}